\documentclass[prb,preprint]{revtex4}

\begin{document}


\title{Pulsed versus DC I-V characteristics of resistive manganites}

\author{B. Fisher, J. Genossar, K. B. Chashka, L. Patlagan, and G. M. Reisner}
\affiliation{Physics Department and Crown Center for
Superconductivity, Technion, Haifa 32000, Israel.}

\date{\today}

\begin{abstract}
We report on pulsed and DC I-V characteristics of polycrystalline
samples of three charge-ordered manganites,
Pr$_{2/3}$Ca$_{1/3}$MnO$_3$, Pr$_{1/2}$Ca$_{1/2}$MnO$_3$,
Bi$_{1/2}$Sr$_{1/2}$MnO$_3$ and of a double-perovskite
Sr$_2$MnReO$_6$, in a temperature range where their ohmic
resistivity obeys the Efros-Shklovskii variable range hopping
relation. For {\it all} samples, the DC I(V) exhibits at high
currents negative differential resistance and hysteresis, which
mask a perfectly ohmic or a moderately nonohmic conductivity
obtained by pulsed measurements. This demonstrates that the widely
used DC I-V measurements are usually misleading.
\end{abstract}


\maketitle

Charge ordering (CO), often associated with spin
(antiferromagnetic (AF)) ordering{\cite{raorev}} and its collapse
in manganite perovskites,{\cite{tomioka}}  have been the subject
of intense investigations over the past decade. The electric field
induced current switching in a CO Pr$_{0.7}$Ca$_{0.3}$MnO$_3$
single crystal, at temperatures below the CO transition
temperature ($T_{CO}$),{\cite{asamitsu}} attracted  great
interest. Investigations of the non-linear conductivity of single
crystals,
 films and polycrystalline samples of different CO manganites
were reported by many groups and gave rise to several {\it ad-hoc}
 interpretations. (For a short review see Ref.
{\onlinecite{monod1}}). {\it All} these I-V characteristics were
obtained with DC currents. In most studies, Joule heating was
ruled out by calculating the expected temperature increment
($\Delta T$) and/or by monitoring it close to the samples'
surface. However, both the calculations and the measurements
showing $\Delta T$ to be negligible were misleading, being based
on the assumption that the current flow throughout the crossection
of the sample and thus its temperature are homogeneous. An
elaborate investigation on polycrystalline samples of
Pr$_{0.8}$Ca$_{0.2}$MnO$_3$ (with no CO but exhibiting nonlinear
conductivity)  has shown that the internal thermal gradient caused
by Joule heating is  the origin of the I(V)
nonlinearity.{\cite{monod1}} Thermal filaments in switching
samples of other materials have been visually observed  in the
past.{\cite{weinrauch}}

An old but lately neglected technique for preventing and
confirming absence of Joule heating errors in high E-field
measurements, consists of applying pulsed currents and following
the time dependence of the response on an oscilloscope. For short
rise-time square pulses, Joule heating is negligible as long as
the response is independent of time.

Here we report on a comparative study of DC and pulsed I-V
characteristics of polycrystalline samples of
Pr$_{1-x}$Ca$_x$MnO$_3$ with $x=1/3$ (PCMO(1/3)) and $x=1/2$
(PCMO(1/2)), Bi$_{1/2}$Sr$_{1/2}$MnO$_3$ (BSMO(1/2)) and
Sr$_2$MnReO$_6$ (SMRO).
 PCMO(1/3), PCMO(1/2), and BSMO(1/2) are CO-AF  manganites
 ($T_{CO}$=240, 220 and 450 K and $T_N$=180, 150 and 150
K, respectively){\cite{tomioka,hejtmanek}} while SMRO is a
double-perovskite that undergoes a transition to insulating
 ferrimagnet at $T_N$=120 K{\cite{popov}}. The resistivity of all
these systems follows, over wide ranges of $T$, the
Efros-Shklovskii variable range hopping (ES-VRH)
relation:{\cite{efros}} $\rho(T) \propto exp(T_o/T)^{1/2}$. We
show that the pulsed $J$ -$E$ characteristics ($J$- current
density and $E$ - electric field)
  are linear for most samples or weakly nonlinear for a few,
 up to fields around 10$^3$ V/cm, while those obtained with DC measurements are nonlinear
 and exhibit negative differential resistance (NDR) at much lower fields.

The  samples were  prepared according to protocols taken from
Ref.{\onlinecite{fisher1}} for PCMO, Ref.{\onlinecite{monoz}} for
BSMO and Ref. {\onlinecite{popov}} for SMRO. X-ray diffraction
(XRD) measurements have been performed using a Siemens D5000
powder diffractometer with CuK$\alpha$ radiation. None of the  XRD
powder patterns showed foreign phases.

Most I-V characteristics were measured by the four-probe technique
(1 and 4 -the current contacts, 2 and 3 -the voltage probes). The
voltage probe separations ranged between 0.9 to 1.6 mm, about 1/3
of the total lengths of the samples. To extend the pulsed
measurements to higher fields and lower temperatures we used also
short samples of about 1 mm length. They were used only when the
two-probe $\rho(T)$,  agreed with the four-probe one (within the
uncertainty of the dimensions). The agreement was very good for
the BSMO(1/2) and SMRO samples at all temperatures and for PCMO
samples below $\sim 100$ K.

DC I-V characteristics were measured over many orders of magnitude
up to the current runaway in the NDR regime. Under this
restriction, the measured surface {\it $\Delta T$ never exceeded a
fraction of a degree}.

For pulsed measurements, single pulses in the millisecond range
were applied from a Keithley 237 high voltage source; its maximal
voltage was 150V for currents up to 10mA. Using  the two channels
of a Tektronix 2221A digital storage oscilloscope we recorded
simultaneously, as function of time the voltage drops between the
ground and pairs of voltage probes. For each pair ( $V_{02}$,
$V_{03}$  and  $V_{01}$, $V_{04}$) the measurement was repeated
four times in order to average out the small fluctuations.
$V_{01}$, on a small resistance in series with the sample and
$V_{14}=V_{04}-V_{01}$ were used  to check the current and the
quality of the current contacts, respectively. A $V_{14}/V_{23}$
ratio, independent of $E$ and $T$ and close to that expected from
the dimensions, is an additional test for the use of two-probe
samples.

Fig. 1 shows $\rho(T)$ of PCMO(1/2), PCMO(1/3), BSMO(1/2) and SMRO
plotted on a semilog graph as function of $T^{-1/2}$. The solid
lines represent the relation $ln\rho =\rho_o exp (T_o/T)^{1/2}$
that fits very well the data  over at least four orders of
magnitude. Unexpectedly, the plots for BSMO(1/2) and SMRO and
sections of the PCMO(1/3) and PCMO(1/2) plots are almost parallel
straight lines. Their slopes correspond to values of $T_o$ within
the range between 5.1$\times10^4$ for PCMO(1/2) to
5.7$\times10^4$, for BSMO(1/2). According to the ES-VRH model,
 $\epsilon_r a=2.8e^2/(4\pi \epsilon_o k_BT_o)$, where
$a$ is the localization length and $\epsilon_r$ - the relative
dielectric constant. The above values of $T_o$ correspond to
$\epsilon_r a \approx 0.9$nm; an atomic length-scale of  $a$,
seems reasonable.

Fig. 2(a) shows the pulsed (symbols) and the DC (solid lines)
$J$-$E$ characteristics measured at various temperatures on
PCMO(1/3), down to 80 K on a four-probe sample, and at 60 and 70 K
on a two-probe one.   The pulsed four-probe $J$-$E$
characteristics are ohmic up to $E \sim 800$ V/cm while those
obtained for the two-probe sample are non-ohmic. For both
two-probe plots $J/J_{ohm}$ ($J_{ohm}$- the extrapolated ohmic
current) is 1.85 at the maximal field ($E= 1480$ V/cm). In Fig.
2(b) the pulsed two-probe $J$-$E$ characteristics of a PCMO(1/2)
sample are perfectly ohmic.

The DC $J(E)$ plots in Fig. 2, for both PCMO(1/3) and PCMO(1/2),
are linear over several orders of magnitude of $E$ but become
strongly nonlinear,    exhibit NDR and hysteresis  at moderate
fields (see insets for better resolution).  The field at the
 turnover into the NDR regime increases with decreasing $T$.
 The main features of these DC characteristics are as calculated
by models of self-heating.{\cite{kroll}}

Fig. 3 shows the pulsed and the DC $J$-$E$ characteristics of
BSMO(1/2) samples, measured at various temperatures, down to 90 K
 on a four-probe sample and for 80 K on a two-probe sample. All
pulsed $J$-$E$ characteristics are ohmic, while the DC $J(E)$ are
similar to those of the PCMO samples.

Fig. 4 shows the pulsed and the DC $J$-$E$ characteristics
measured  on SMRO samples, down to 120 K on a four-probe sample,
and at 100 K, on a two-probe sample.  The pulsed $J$-$E$
characteristics
 are nonlinear at  fields as low as  $\sim$ 100 V/cm.
The non-linearity increases with decreasing $T$ and increasing
$E$.  DC and pulsed $J(E)$ are identical in the low-field
nonlinear regime  but diverge at higher fields. For the pulsed
measurements, $J/J_{ohm}$ is 2.7 at $T=120$ K and $E=580$ V/cm
and, 4.6 at $T=100$ K and $E=1560$ V/cm. After separation, the DC
$J(E)$ increase much faster towards the NDR regime. The
nonlinearity in the pulsed $J(E)$ data for both types of SMRO
samples, indicates that in the case of PCMO(3) two-probe sample it
may not be an artifact of this  configuration. Nonlinear $J(E)$
 determined by the parameter $eEa/k_BT$ is expected in case of hopping conductivity.{\cite{pollak}}
 For $T\geq 100$ K and $a\approx 1nm$ (see above), one would expect nonlinear
 $J(E)$ at much higher fields than reported here.
On the other hand, the various models suggested for the nonlinear
conductivity in CO manganites are irrelevant for SMRO.

The absence of broad band and telegraph noise that would be
associated with CO depinning, filament formation or other
instabilities is confirmed by the flattness (within three digit
resolution, over several milliseconds) of the observed voltage
drops.

In summary, for {\it all} samples investigated here, CO manganites
and a resistive double-perovskite,  the DC I-V characteristics
mask a perfect linearity or a moderate nonlinearity of $J(E)$
observed by pulsed measurements. This demonstrates that the widely
used DC I-V measurements are usually misleading.

%

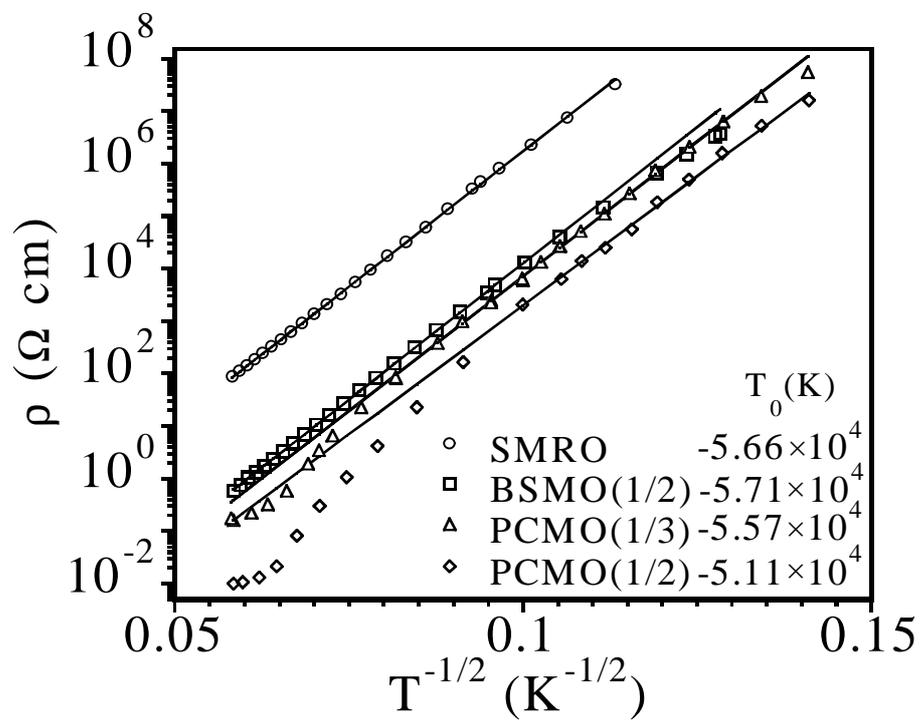
\begin{figure}[p]
\caption{Semilog plot of $\rho$ versus $T^{-1/2}$ for
Pr$_{2/3}$Ca$_{1/3}$MnO$_3$, Pr$_{1/2}$Ca$_{1/2}$MnO$_3$,
Bi$_{1/2}$Sr$_{1/2}$MnO$_3$, Sr$_2$MnReO$_6$. Solid lines
represent the relation $\rho=\rho_oexp(T_o/T)^{1/2}$ that fits
each set of experimental data over at least four orders of
magnitude of $\rho$.  }
 \label{1}
 \end{figure}

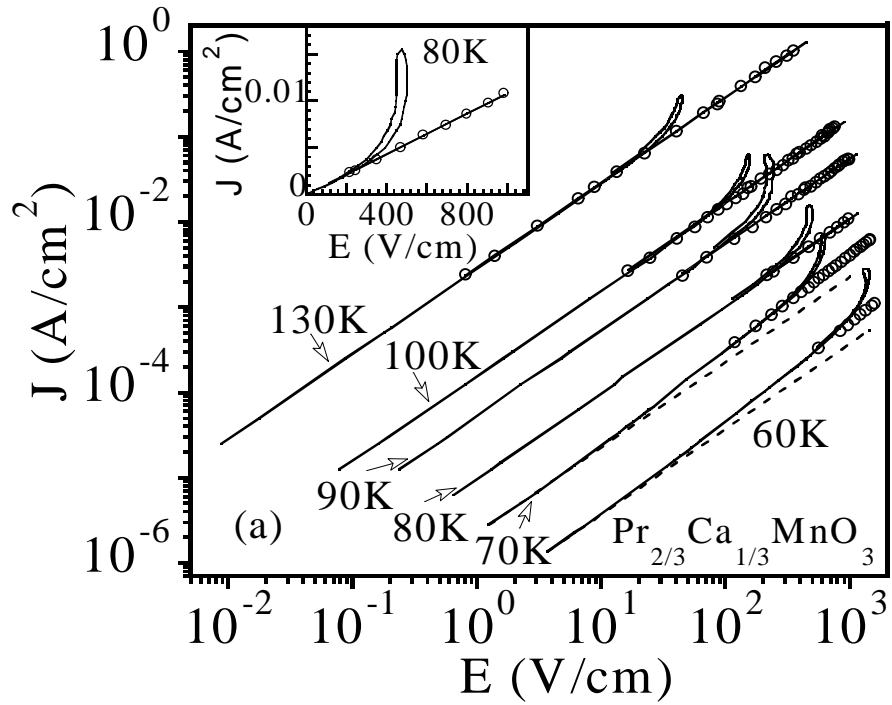
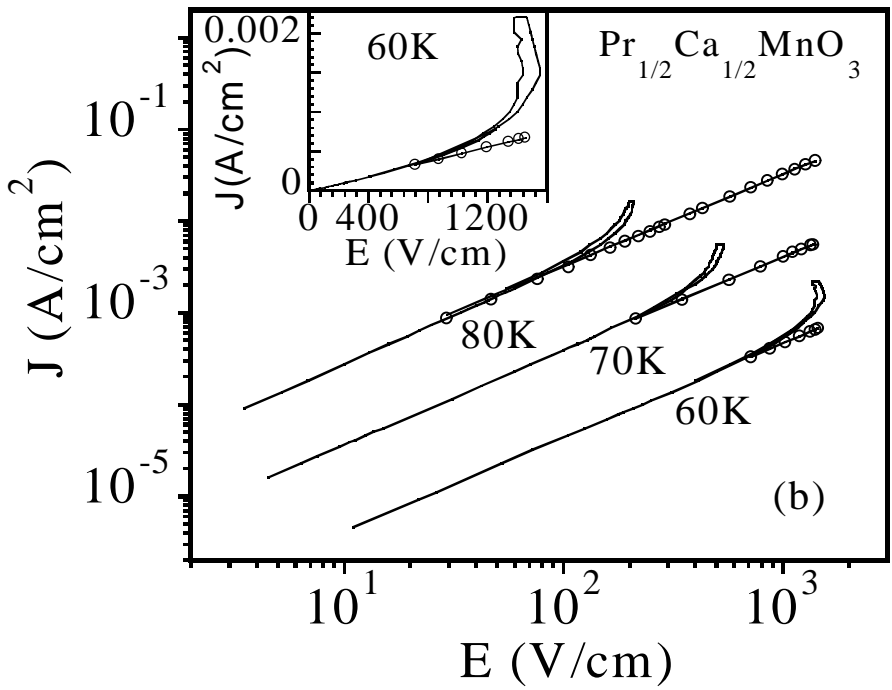
 \begin{figure}[p]
\caption{$J$-$E$ characteristics  of (a)
Pr$_{2/3}$Ca$_{1/3}$MnO$_3$ samples
 measured in four-probe, and two-probe
configurations, down to 90 K and below, respectively
 and (b) on a Pr$_{1/2}$Ca$_{1/2}$MnO$_3$ sample, in a two-probe
 configuration. Symbols for pulses and solid lines for DC. Dashed lines in (a)
represent the extrapolated ohmic current - $J_o$. Insets show
$J(E)$ on linear scale for 80 K in (a) and for 60 K in (b).}
 \label{1}
 \end{figure}

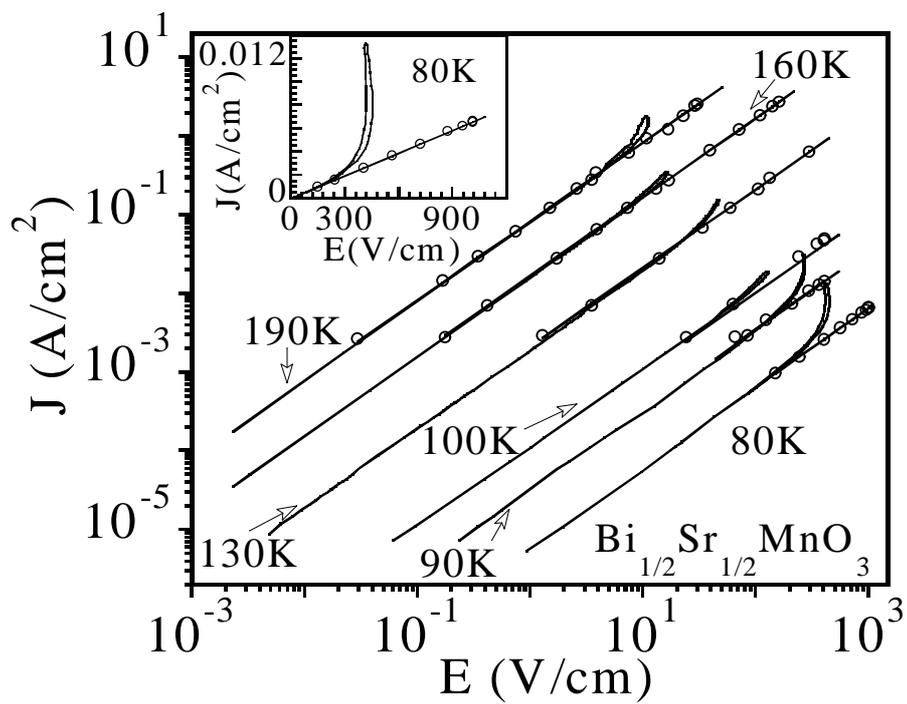
\begin{figure}[p]
\caption{$J$-$E$ characteristics  of Bi$_{1/2}$Sr$_{1/2}$MnO$_3$
samples (in four-probe, or two-probe configurations, down to 90 K
or at 80 K, respectively). Symbols for pulses and solid lines for
DC. The inset shows the plots for 80 K on a linear scale.}
 \label{1}
 \end{figure}

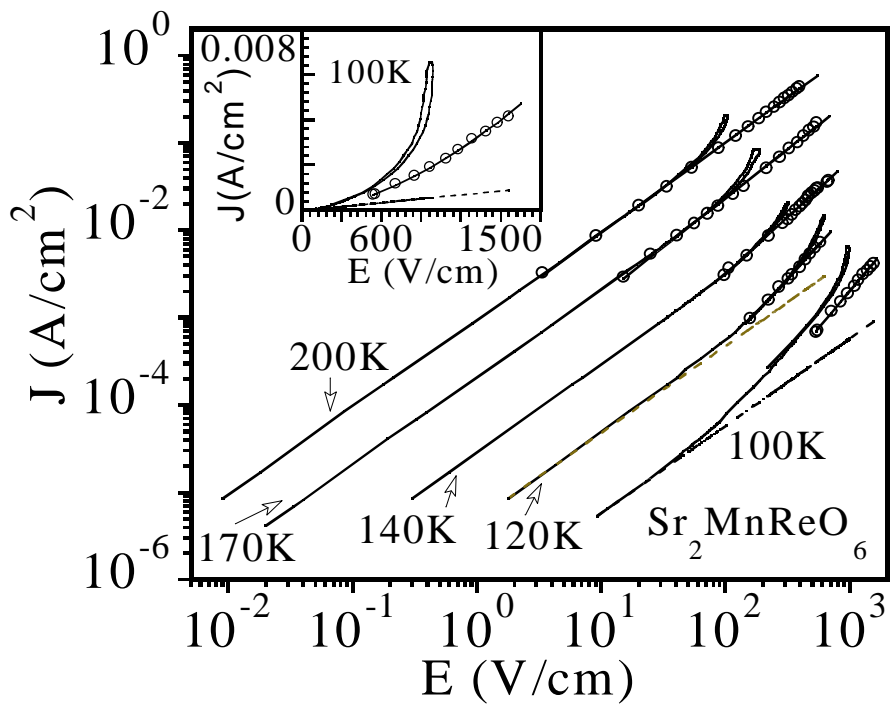
\begin{figure}[p]
\caption{$J$-$E$ characteristics of Sr$_2$MnReO$_6$ samples (in
four-probe, or two- probe configuration, down to 120 K or at 100
K, respectively). Dashed lines represent the ohmic currents.
Symbols for pulses and solid lines for DC. The inset shows the
plots for 100 K on a linear scale. }
 \label{1}
 \end{figure}

\bibliography{basename of prb.bib file}

\end{document}